Lucia Helena Horta Oliveira
EEEM Dr. Silva Mello
Lucia-horta@hotmail.com

**A Escola vai à Mostra de Astronomia do ES: diálogos entre a educação formal, não formal e informal para o desenvolvimento da cultura científica**

RESUMO

Esse artigo derivou do desenvolvimento de atividades de educação em Astronomia, com vistas a incentivar a formação de futuros pesquisadores. Seu objetivo foi apresentar como ocorreu a participação de estudantes da Escola Estadual de Ensino Médio Dr. Silva Mello na Mostra de Astronomia do ES e de como isso promoveu seu envolvimento e participação nas três modalidades de educação, formal, informal e não formal, destacando a articulação entre elas. Com intuito de formar cidadãos com interesses para o desenvolvimento de pesquisas científicas sobre termos astronômicos, desenvolvendo o gosto pela busca de conhecimento e para motivação dessa ação, formamos grupos para apresentação de trabalhos na Mostra de Astronomia do ES, utilizada como espaço de educação não formal. O evento contou com apresentação de trabalhos, palestras e premiações. Como resultado, um dos trabalhos apresentados pelos alunos foi premiado. Ganhou especial relevância a educação informal, momento em que os alunos discutiram sobre ciências, trazendo-a para suas vidas diárias.

**Palavras-chave:** Ensino em Astronomia. Mostra de Astronomia do Espírito Santo. Educação formal, não formal e informal.

ABSTRACT

This article derived from the development of activities on Astronomy education and an incentive to train future researchers, based on High School students. It aimed to present how the students participation from the State High School Dr. Silva Mello took place in the ES Astronomy Exhibition and how it promoted their involvement in the three types of education, formal, informal and non-formal, highlighting the articulation between them. In order to train citizens with interests for the development of scientific research on astronomical terms, developing a taste for the search for knowledge and to motivate this action, we formed groups to present works at the ES Astronomical Exhibition, used as a non-formal educational setting. The event includes presentation of works, lectures, and awards. As a result, one of the works presented by the students was awarded and we realized that informal education gained special relevance, becoming a moment in which students discuss science and bring it into their daily lives.

# 1 Introdução

A Astronomia como ciência que pesquisa os corpos celestes pode ser considerada a primeira ciência já registrada. Desde os persas, hindus e outras culturas orientais, o estudo dos movimentos dos corpos celestes e a aplicação de seus conhecimentos sempre buscaram entender o cotidiano do homem. Até hoje, a Astronomia encanta e desperta interesse. Em um planetário, na escola, ao ar livre, em um sítio arqueológico, a leitura do cosmos motiva. Assim, a aprendizagem de Astronomia e o desenvolvimento pelo gosto científico podem acontecer em diversos espaços, sobre os quais passamos a tratar.

De um modo geral, a diferença entre educação formal, não formal e informal pode ser estabelecida a partir do espaço escolar. As ações educativas escolares seriam as formais e aquelas realizadas fora da escola seriam as não formais e as informais seriam as realizadas no seio familiar, nos grupos de convívio, roda de amigos, onde também se aprende, fora da escola. A educação não formal pode ocorrer em instituições como museus, centros culturais, ONGs e a educação informal ocorre, também, por meio da mídia em geral [6]. Quanto aos espaços, é possível estabelecer a seguinte diferença: o espaço formal é a escola e o espaço não formal, seria qualquer espaço fora da escola, onde uma ação educativa pode ocorrer.

Quando tentamos expressar a forma de tornar conteúdos científicos acessíveis à população, encontramos termos como: difusão, divulgação, disseminação e popularização da ciência. O fato é que, tanto no espaço formal quanto no espaço de educação não formal, caso de um planetário, é possível contribuir para que a ciência chegue ao grande público, capacitando-o a compreender o mundo e o cosmos [6]. Apesar de muitas distinções entre as formas de socializar a ciência, o fim será sempre o mesmo, levar a ciência ao grande público [12]. Desse ponto de vista, Gouvêa [2] defende que o termo popularização da ciência deve ser mais apropriado a ações de divulgação científica feitas diretamente ao público, tornando o termo mais completo que disseminação ou difusão da ciência.

Baseado na revisão bibliográfica apresentada por Marandino et al. [4] e reconhecendo a não existência de uma definição final entre os termos, vez que é grande o elenco de autores que se debruça a pesquisar as questões relativas tanto à educação formal, educação não formal e informal, bem como a potencialidade dos espaços para divulgação científica, neste artigo, apresentaremos uma discussão tanto sobre tais terminologias quanto sobre a finalidade da popularização da ciência.

Assim, o objetivo deste trabalho foi apresentar como ocorreu a participação de estudantes da Escola Estadual de Ensino Médio Dr. Silva Mello na Mostra de Astronomia do ES e de como isso incentivou e promoveu seu envolvimento e participação nas três modalidades de educação, destacando a forte articulação entre elas. Esta escola localiza-se no município de Guarapari e possui dois turnos, com 28 turmas de Ensino Médio, somando um total de quase mil alunos matriculados. Participaram da mostra cinco grupos de alunos do primeiro e três do terceiro ano do Ensino Médio, cada um dos grupos contendo cinco alunos.

## 2 Educação formal como base do aprendizado

A educação formal ocorre em ambiente escolar, onde o currículo oficial ordena o desenvolvimento dos trabalhos. De forma a trazer temas motivadores para o espaço escolar, foram adaptados, nas turmas de primeiro e do terceiro ano, temas gerais da Astronomia ao conteúdo básico lecionado normalmente para o ensino de Física, tais como cosmologia, arqueoastronomia, estações do ano, calendários, o que provocou muito interesse nos alunos. Aproveitamos, quando possível, a interdisciplinaridade com outras matérias, tais como Matemática, Literatura e Geografia. Isso, apesar de não ser muito usual o estudo da Astronomia durante a trajetória formativa do aluno do Ensino Médio, na formação do futuro professor ou no ambiente escolar. A despeito disso, podemos usar a Astronomia como uma articulação para a formação de saberes disciplinares. Essa ponte se torna importante, também, para a facilitação da aprendizagem de Astronomia nos espaços não formais, principalmente para a diminuição de erros conceituais, concepções alternativas, mitos e crenças sobre fenômenos astronômicos. Ao abordar temas como origem do Sol, da Lua, o conceito de calendário, as diferenças entre solstício e equinócio, contamos com a participação dos professores de Literatura, apontando esses eventos nas culturas antigas, do professor de Geografia e do professor de Matemática, com relação aos cálculos de distância entre o planeta Terra e outros planetas do sistema solar.

A educação formal é a base de que precisamos para dar o pontapé inicial para que os trabalhos se desenvolvam e possam ocorrer nos espaços não formais. Sem a educação formal fica mais difícil esses trabalhos alcançarem cunho científico. Isso é importante para que se desenvolva o espírito científico nos alunos, para que eles consigam superar a curiosidade ingênua passando à curiosidade científica (FREIRE, 2011) [1]. Por exemplo, pelo fato de a escola estar localizada no município de Guarapari, consta do imaginário dos alunos a influência da Lua sobre as marés, mas, até então, eles não tinham estudado esse tópico do ponto de vista científico. Assim, eles podem se aprimorar, quando surgirem temas tais como a Lua, as marés, as estações do ano, o movimento da Terra na educação informal. Em diálogos e trocas, eles poderão discutir com propriedade esses assuntos.

Neste artigo, discutiremos as contribuições da Mostra de Astronomia do Espírito Santo promovida pela UFES para conhecimento de temas como Astrofísica, Astronomia, Astronáutica e Cosmologia, tomando a mostra como espaço de educação não formal. A participação na Mostra partiu de uma iniciativa da autora deste artigo que, na época, era aluna do Mestrado Profissional em Ensino de Física do IFES. O desenvolvimento dos trabalhos apresentados na Mostra foi enriquecido pela orientação de professores da UFES, o que tornou possível manter uma parceria entre Universidade Federal e escolas públicas estaduais.

Nesse evento anual, os alunos apresentam trabalhos e competem com trabalhos de outras instituições. A participação na Mostra movimentou o espaço escolar, trazendo para o centro do palco o protagonismo dos alunos e a contribuição do corpo técnico e pedagógico.

**3 A Mostra de Astronomia do ES (MAES) como espaço de educação não formal**

O segundo aspecto a ser analisado é a educação não formal e o espaço onde esta ocorre. A educação não formal tem caráter sempre coletivo e envolve práticas educativas desenvolvidas fora da escola, em espaços previamente selecionados para tal, sem a obrigatoriedade de seguir os ritos do espaço escolar. No espaço não formal, quebram-se hierarquias, superam-se medos, desenvolvem-se potencialidades. Nesse caso em específico, os alunos acompanharam a professora a locais fora da escola, tais como, morros e praias no município de Guarapari, ES, onde puderam montar os telescópios adquiridos pela EEEM Dr. Silva Mello, aprendendo, na prática, sua montagem, instrumentação, apontamento e observação do céu. Nessas atividades em espaços não formais, os alunos trazem um planejamento de estudos para utilizarem onde poderão observar, estudando as características e a utilização dos equipamentos, suas curiosidades, mapeando a abóbada celeste localizando planetas, estrelas, constelações etc... Nesse contexto trabalham como transmissores do conhecimento aprendido no espaço formal. Nos espaços fora da escola utilizados para observação do céu, os alunos recebem muitos visitantes que interagem e dialogam, fazendo-os assumir a condução do trabalho.

Essas ações foram orientadas pela professora e contavam com o apoio do Instituto Federal do Espírito Santo, considerando que em 2018 cursava o mestrado em Ensino de Física desta instituição, e alguns tópicos foram discutidos com os professores do mestrado, tendo em vista a preparação dos alunos. Em outras ocasiões, as ações foram apoiadas pela Universidade Federal do Espírito Santo, por meio do Programa de Pós-Graduação em Astrofísica, Cosmologia e Gravitação da UFES (PPGCosmo) e do Núcleo de Astrofísica e Cosmologia da UFES (Núcleo Cosmo-Ufes) [17], no que diz respeito à utilização do material de apoio, com empréstimo de telescópio, palestras e orientações.

A presença das entidades de educação superior, no processo de ensino-aprendizagem em espaço não formal, é feito através de orientações e palestras. O sucesso do projeto se dá em decorrência da união das forças em favor da iniciação científica e de atividades de popularização da ciência. Queiroz *et al.* (2002) [9] demonstram que a educação não formal utilizada em espaço não formal institucionalizado tem características próprias e diferentes da educação formal, feita dentro da escola.

Contudo, ao usarmos esses espaços não formais, precisamos ficar atentos à não escolarização deles (Gouvêa *et al.,* 1993 [3]; Maradino, 2000[5], Marandino et al.,2004[4]; Queiroz *et al*., 2002[9]; Vieira, 2005[12]; Pivelli, 2006 [7]). As práticas escolares não podem ser repetidas nos espaços não formais, sob pena de descaracterizarmos sua riqueza e criatividade. Pivelli e Kawasak (2005, p. 9) [8], alertam:

> [...] é preciso ter cuidado para não escolarizar as instituições. Acredita-se que o objetivo maior destes locais que expõem biodiversidade é o de

> despertar curiosidades, paixões, possibilitar situações investigadoras, gerar perguntas que proporcionem a sua evolução e não somente respostas às questões que são colocadas pelo ensino formal (PIVELLI; KAWASAK, 2005, p. 9).[8]

Rocha e Fachim-Terám (2010) [10] destacam a relevância da escola no processo, ao discutirem a importância dos espaços não formais para o ensino de Ciências, e acentuam a impossibilidade de alcançar uma educação científica mais completa sem a parceria da escola com esses espaços. Assim, os espaços não formais devem atuar em complementaridade aos espaços formais. Nesses espaços, os alunos pesquisam, aprendem, se planejam e ensinam, experimentando a iniciação científica. A educação não formal se transforma em um observatório móvel e acessível a toda a população.

Além das incursões aonde a população tem acesso fácil, o grupo de alunos da referida escola geralmente participa de atividades específicas com a orientação de professores doutores da UFES e do Instituto Federal do Espirito Santo, tais como visitas, seminários e minicursos. Essas incursões geralmente saem da rotina escolar e os alunos têm contato com um aprendizado mais científico, muitas vezes aprendendo cálculos sofisticados específicos da Astronomia, como, por exemplo, cálculos da Astronomia de Posição, ensinados por especialistas da área.

Outra atividade da educação não formal realizada pelo grupo de alunos foi ajudar na manutenção e no trabalho de estudo da estação de vídeo monitoramento de pequenos objetos, a qual foi adquirida pelo projeto do Grupo de Astronomia de Guarapari – Dr. Silva Mello (GAG DSM), formado em 2018, por cerca de 50 alunos desta mesma escola, dos quais três são orientandos da autora deste artigo e receberam bolsa de iniciação científica em 2018, provenientes do CNPq. Esse projeto conta com uma câmera com lente especial que monitora uma área específica do céu noturno e trabalha na suíte UFO Capture e Analyzer, em parceria com o projeto Exoss Citizen Science [13]. O processo se completa com o pareamento com outra estação, para obter confiabilidade nos dados de órbitas dos objetos registrados (meteoros). Os alunos trabalharam com ele, durante o ano de 2019, conseguindo desenvolver pesquisas interessantes para seus estudos. Esse projeto foi responsável por desenvolver uma pesquisa sobre estudos de captação dos meteoros com procedência da constelação de Hydra, o cálculo provável de suas velocidades, a magnitude média, a quantidade de capturas feitas pela câmera e o período dessas capturas.

A pesquisa que foi apresentada na última MAES (2019) classificou o grupo de estudos em primeiro lugar em sua categoria, o de grupo de estudantes de Escolas Estaduais. Com o projeto em andamento, a escola conseguiu a doação de mais uma câmera feita pelo projeto Exoss Citizen Science, que funciona com um programa novo, a princípio instalado como teste. O programa foi desenvolvido pela NASA [22] e tem a EXOSS [19] como correspondente no Brasil. O algoritmo CAMS é mais moderno e preciso e, com o pareamento, ele atinge grau ótimo de confiabilidade.

O projeto iniciou-se com parceria de análises feitas do telescópio Pan Starrs [25], que conta com o *software* Astrométrica [26], no ano de 2018 e que é bem conhecido de escolas nacionais e internacionais. O programa Astrométrica possui um software de acesso livre. Através de cadastro prévio, disponibiliza para escolas e grupos de Astronomia do mundo inteiro 4 imagens sequentes, mensais, tiradas pelo telescópio Pan Starrs que fica no Havaí. Através dessas imagens sobrepostas, os voluntários cadastrados procuram vestígios de movimentações parecidas com objetos que podem ser classificados como asteroides. Os objetos localizados são marcados e enviados através de um código em um relatório para uma área de pesquisa da NASA. Os colaboradores que encontram esses objetos recebem um certificado de participação no programa. Posteriormente esse objeto estudado pode ser confirmado como um asteroide. Nesse caso, o colaborador que enviou o código do objeto encontrado tem o direito de colocar o nome que quiser no objeto. Essa é uma parceria que colabora para a busca de possíveis asteroides e é de cunho internacional.

A partir do trabalho envolvido nesse projeto, um grupo de alunos apresentou estudos desenvolvidos sobre o funcionamento do *software* Astrométrica na primeira Mostra Estadual de Astronomia, Astrofísica, Astronáutica e Cosmologia em 2018, ficando como finalista, recebendo três bolsas de iniciação científica júnior premiadas no evento.

Com a devida orientação do Dr. José Antônio de Freitas Pacheco (*Observatoire de la Côte d' Azur*, França) [23], o projeto tomou um rumo próprio, desenvolvendo pesquisas com parceria do Observatório Nacional [18] e da rede Exoss [19].

Hoje, os alunos fazem pesquisas e colaborações nacionais e internacionais. Entre as colaborações nacionais temos: Observatório Nacional (ON) Projeto IMPACTUN [18], a rede EXOSS [19], União Brasileira de Astronomia (UBA) [21], Sociedade Brasileira de Astronomia (SAB) [20], IFES Campus Guarapari, PPGCOSMO-UFES e o Núcleo Cosmo-UFES [17], entre os colaboradores internacionais temos a NASA, através da EXOSS CAMS [22], o projeto Neoshield 2 [14], com a divulgação do Asteroid Day [15] nas escolas e o projeto Garatéa-ISS [16].

Subjacente a essas colaborações, existe uma parceria público-privada com a NASA, que busca formas alternativas para proteger a Terra de uma possível extinção em massa.

Esse trabalho é feito por uma equipe de diferentes lideranças mundiais (IMPACTUM e NASA, através da EXOSS CAMS, e cientistas da NEOShield 2) que estuda, e propõe, soluções diferentes para prevenção de uma ameaça teórica onde são feitos estudos e simulações. O objetivo é criar estratégias para eliminação de possíveis ameaças celestes de tamanhos elevados, que possam invadir nossa atmosfera, como asteroides e cometas. Mesmo com pequenas chances de que a Terra possa ser atingida por corpos celestes gigantes, as ameaças são reais.

O projeto "Estamos de Olho no Céu" desenvolvido na EEEM Dr. Silva Mello, colabora com essa iniciativa por meio do monitoramento de certa parte do céu noturno, analisando as imagens e filtrando o que for necessário para os devidos estudos. Nas imagens captadas do céu noturno, coberto pela câmera, são flagrados desde vaga-lumes a aviões, que, às vezes, têm características de meteoros nas imagens que aparecem durante o vídeo-monitoramento. Esse ruído é considerado falso-positivo e deve ser retirado, manualmente, antes de ser enviado para pesquisas mais aprofundadas.

São diversas imagens geradas todos os dias e filtrar essas informações é muito importante para futuras descobertas dos pesquisadores, que têm as imagens e as informações mais rapidamente e mais precisas para realizar seus estudos com segurança.

Os dados espaciais costumam ser massivos, multidimensionais, difíceis e quase impossíveis de analisar por completo. Entre os estudos que são feitos pelos cientistas da Neoshields 2 temos: modelagem do formato de NEOs (Near Earth Objects), aprender mais sobre os cometas misteriosos de longos períodos, traçar a modelagem de suas rotas, observação e estudos de novas chuvas de meteoros (radiantes), mapeamento de superfície de objetos celestes, com o intuito de encontrar água e minerais, buscando formas de utilizar essa água para possíveis voos espaciais a distâncias maiores[24].

**4 Educação informal como consequência**

Como terceiro aspecto, temos a educação informal, que não possui intencionalidade, pois é decorrente de momentos não organizados e espontâneos, do dia a dia, durante a interação com familiares amigos e conversas ocasionais. A esse respeito surge a ciência durante a observação, na instrução do curioso, da orientação familiar. A informação científica se torna normal na vida da maioria dos alunos que experimenta Astronomia no espaço formal e não formal. O espaço informal é consequência do aprendizado e a multiplicação do conhecimento como um hobby e está diretamente ligada à popularização da ciência. Podemos determinar que o objetivo vai além da divulgação, pois são consideradas as necessidades e as expectativas do público-alvo, buscando a curiosidade cultural vasta da Astronomia ao divulgar o conhecimento à população. Isso traz uma satisfação própria ao aluno que sente a importância da sua atitude no processo como o centro da comunicação e do ensino naquele momento. Essa divulgação apresenta sempre resultados muito positivos para a segurança na interlocução do aluno, no desenvolvimento da pesquisa e do aprendizado.

Segundo Marandino [5], pesquisas na área de ensino de Ciências em espaços não formais são muito escassas, portanto temos poucas referências para nos baseamos. Por esse motivo os resultados obtidos se tornam muito interessantes para os próximos pesquisadores.

O apoio à utilização pedagógica dos espaços não formais e à educação informal atua diretamente no desenvolvimento da cultura científica, momento em que os alunos utilizam, como assunto para os diálogos e as trocas, a ciência aprendida na escola. Alunos que se envolvem nas atividades não formais aprendem a intensificar a necessidade da educação formal, por se tornarem responsáveis por informações confiáveis quando ocorre a informalidade.

> Nesse sentido é importante salientar que o campo de educação não escolar (informal e não formal) sempre existiu com o campo de educação escolar, sendo mesmo possível imaginar sinergias pedagógicas muito produtivas e constatar experiências com intersecções e complementariedades várias (AFONSO *apud* SIMSON; PARK; FERNANDES, 2001, p. 31) [11].

Percebemos que, quando os alunos passam as informações, como nas incursões de conscientização do Asteroid Day [15], a outros alunos em outras escolas, o nível de atenção dos ouvintes é bem maior do que se fosse o mesmo assunto abordado em uma palestra proferida por um professor. Geralmente os alunos ouvintes se projetam nos alunos que estão dando as palestras, assim absorvendo muito mais a informação e transmitindo depois informalmente em sua comunidade. Com relação à participação dos alunos na Mostra de Astronomia do Espírito Santo, eles foram preparados para explorar ao máximo as atividades no espaço não formal, para que, ao mesmo tempo, aprendessem, transmitissem o conhecimento adquirido ensinando aos curiosos que sempre aparecem.

A Escola Doutor Silva Mello realiza pelo menos cinco incursões em escolas da região por ano, levando informações sobre corpos celestes, conscientizando a população sobre os perigos que podem vir do céu e as ações que temos feito para monitorar e proteger a Terra desses corpos celestes. Em algumas ocasiões, já foram organizadas aulas de campo para que os alunos se reunissem em acampamentos, que ficam em locais com altitude mais adequada, para estudos estelares, para aprimoramento do conhecimento adquirido. Esses momentos são ricos de aprendizagem científica, onde eles têm o conhecimento apresentado por cientistas, ajudando no fortalecimento da vontade ao estudo, pois eles podem interagir e tirar suas dúvidas com especialistas dos mais diversos assuntos do universo. Com essa forma de transmissão de conhecimento, utiliza-se de criatividade para a palestra informal e criam-se formas de tornar essa informalidade em gosto pelo conhecimento e pela ciência.

**5 Considerações finais**

A utilização do espaço formal geralmente funciona como base para a aquisição do conhecimento. Nesse trabalho, ele funcionou como base para desenvolvimento de atividades em espaços não formais. A aprendizagem em espaços não formais gerou desenvolvimento de pesquisas e atuou em complementaridade à educação formal para aplicação nos projetos desenvolvidos. Diante desse cenário, temos a educação informal como consequência lógica do aprendizado vivenciado, como consequência da participação dos estudantes nas atividades formais e não formais. O apoio ao não formalismo contribui diretamente para aumentar o interesse pela educação formal e informal dos alunos. Alunos que se envolvem nas atividades não formais aprendem a intensificar a necessidade da educação formal, por se tornarem responsáveis por informações confiáveis, quando ocorre a informalidade.

A participação dos alunos nas apresentações de trabalhos na Mostra de Astronomia do Espírito Santo, sendo usada como espaço de educação não formal para aprendizagem de temas de Astronomia, além de envolver a comunidade escolar e promover a interação entre participantes, funcionou como momentos para aprendizagem colaborativa, criando uma rede de conhecimento e pesquisa científica que cresce com novas atividades e novos personagens a cada ano. Os alunos se organizam e mantêm pesquisas e estudos focados nas próximas edições. A MAES, ao ser utilizada como ponte de aprendizado, criou uma vasta célula de multiplicação de saberes aos alunos, o que torna a experiência muito importante na concretização do aprendizado formal exigido no ambiente escolar.

**Agradecimentos:**

*Sobre a autora*
Lúcia Helena Horta de Oliveira (luciahorta@hotmail.com) tem mestrado em ensino de física pelo IFES (2020) e é professora da EEEM Dr. Silva Mello (Secretaria de Educação do Estado do Espírito Santo) em Guarapari, ES. É Co-organizadora da Mostra de Astronomia do Espírito Santo.